\documentclass{aastex61}

\newcommand\aastex{AAS\TeX}

\received{}
\revised{}
\accepted{}
\submitjournal{ApJ}

\shorttitle{\aastex\ On the Impact Origin of Phobos and Deimos I}
\shortauthors{Hyodo et al.}

\begin{document}

\title{On the Impact Origin of Phobos and Deimos I:\\Thermodynamic and Physical Aspects}

\correspondingauthor{Ryuki Hyodo}
\email{hyodo@elsi.jp}

\author{Ryuki Hyodo}
\affil{Earth-Life Science Institute/Tokyo Institute of Technology, 2-12-1 Tokyo, Japan}
\affiliation{Institut de Physique du Globe/Universit{\'e} Paris Diderot/CEA/CNRS, 75005 Paris, France}

\author{Hidenori Genda}
\affil{Earth-Life Science Institute/Tokyo Institute of Technology, 2-12-1 Tokyo, Japan}

\author{S{\'e}bastien Charnoz}
\affiliation{Institut de Physique du Globe/Universit{\'e} Paris Diderot/CEA/CNRS, 75005 Paris, France}

\author{Pascal Rosenblatt}
\affiliation{Royal Observatory of Belgium, 1180 Brussels, Belgium}



\begin{abstract}
Phobos and Deimos are the two small moons of Mars. Recent works have shown that they can accrete within an impact-generated disk. However, the detailed structure and initial thermodynamic properties of the disk are poorly understood. In this paper, we perform high-resolution SPH simulations of the Martian moon-forming giant impact that can also form the Borealis basin. This giant impact heats up the disk material (around $\sim 2000$ K in temperature) with an entropy increase of $\sim 1500$ J K$^{-1}$ kg$^{-1}$. Thus, the disk material should be mostly molten, though a tiny fraction of disk material ($< 5\%$) would even experience vaporization. Typically, a piece of molten disk material is estimated to be meter sized due to the fragmentation regulated by their shear velocity and surface tension during the impact process. The disk materials initially have highly eccentric orbits ($e \sim 0.6-0.9$) and successive collisions between meter-sized fragments at high impact velocity ($\sim 3-5$ km s$^{-1}$) can grind them down to $\sim100 \mu$m-sized particles. On the other hand, a tiny amount of vaporized disk material condenses into $\sim 0.1\mu$m-sized grains. Thus, the building blocks of the Martian moons are expected to be a mixture of these different sized particles from meter-sized down to $\sim 100 \mu$m-sized particles and $\sim 0.1\mu$m-sized grains. Our simulations also suggest that the building blocks of Phobos and Deimos contain both impactor and Martian materials (at least 35\%), most of which come from the Martian mantle (50-150 km in depth; at least 50\%). Our results will give useful information for planning a future sample return mission to Martian moons, such as JAXA's MMX (Martian Moons eXploration) mission.
\end{abstract}




\section{Introduction} \label{sec:intro}
The origin of the two small moons, Phobos and Deimos, is still debated. It has been believed that they were captured asteroids due to their spectral properties sharing a resemblance with D-type asteroids \citep[e.g.][]{Mur91,Bur78}. However, the captured scenario is confronted with the difficulty of explaining their almost circular equatorial orbits around Mars \citep{Bur92}. In contrast, accretion within an impact-generated disk may naturally explain their orbital configurations \citep{Ida97,Hyo15,Hyo15b,Hyo17b}. \cite{Cra94,Cra11} have proposed that Phobos and Deimos may form within a debris disk that formed by a giant impact. Then, \cite{Ros16} have shown that these small moons can be formed by accretion within a thin debris disk that extends outside the Roche limit and is sculpted by the outward migration of a large inner moon that is formed by the spreading of a thick disk lying inside the Roche limit. Furthermore, \cite{Hes17} have shown that tidal disruption of such a large inner moon during the tidal decay creates a new generation of rings around Mars followed by the spreading and accretion of smaller moons. They showed that such a process could occur repeatedly over the past 4.3 billion years, suggesting that Phobos is the last generation of moon we observe today.\\

As discussed above, the accretion within a debris disk can explain the masses and orbital characteristics of Martian moons \citep{Ros16,Hes17}. In the case of \cite{Ros16}, Phobos and Deimos accrete in a low-mass extended outer disk that resides between $4-7 R_{\rm Mars}$, where $R_{\rm Mars}$ is the Martian radius. So, the building blocks of both Phobos and Deimos are initially distributed in this outer region. In contrast, following the \cite{Hes17} scenario, Phobos forms from the spreading of a Roche-interior disk after several cycles of tidal destructions and accretions of ancient larger moons; that is, the building blocks of Phobos originally come from inside the Roche limit ($\sim 3R_{\rm Mars}$). Note that, their model explained the formation of Phobos, but not that of Deimos. Currently, the only successful scenario to form Deimos is the accretion within an extended outer disk proposed by \cite{Ros16}.\\

A debris disk around Mars is formed by a giant impact that have created the Borealis basin, the asymmetric northern lowland \citep{Mar08}. The canonical Borealis-forming impactor has a mass of about $0.03$ Mars mass (about 1/3 Mars radius) and an impact velocity of $\sim 6$ km s$^{-1}$ \citep{Mar08, Cit15, Ros16}. However, in the previous papers, due to lack of resolution in simulations, the detailed disk structure is unclear. In addition, due to the limitation of the equation of state that was used, the thermodynamic properties of the disk are not well characterized. Such information is important for bridging the gap between dynamical models and the observed properties of the Martian moons.\\

Now JAXA (Japan Aerospace eXploration Agency) is planning the Martian Moons eXplorer (MMX) mission, in which they will send a spacecraft to Martian moon(s) and get samples there. If accretion scenarios discussed in \cite{Ros16} or \cite{Hes17} are correct, the debris formed by a giant impact are the building blocks of Martian satellites. In addition, the Martian moon-forming impact can blast some Martian material into orbits around Mars as debris. Therefore, we may have a chance to collect not only samples from the impactor, but also from Mars. Thus, detailed studies of the impact and formation of the disk is now important to constrain the expected compositions as well as physical properties of the initial disk particles in order to maximize the scientific value of the MMX mission.\\

In this paper, using high-resolution smoothed particle hydrodynamics (SPH) simulations, we investigate the canonical Martian moon-forming giant impact and constrain the structure and the thermodynamic properties of the initial debris disk. In section 2, we describe our numerical method and models. In section 3, we present our numerical results and discuss the disk properties resulting from the canonical impact. In section 4, we summarize our results.

\section{Method \& Models}
We applied the SPH method to investigate the giant impact between Mars and an impactor, as in \cite{Cit15} and \cite{Ros16} did. The SPH method is a Lagrangian method \citep{Mon92} in which hydrodynamic equations are solved by considering averaged values of particles through kernel-weighted summation. In this work, we used the GADGET-2 code \citep{Spr05} modified by \cite{Cuk12} to include tabulated equations of state. Regarding the equation of state, M-ANEOS \citep{Mel07} is used to model a differentiated object as the mantle is represented by pure forsterite and the core is represented by pure iron. In this modified version of GADGET-2 \citep{Cuk12}, the entropy equation is used instead of the energy equation. Under this treatment, the entropy only changes through shock propagation, preventing any unphysical change during adiabatic processes. This has strong advantages, especially for achieving accurate temperature and entropy time-evolutions. \cite{Rei17} pointed out that some previous works \citep[e.g.][]{Nak15} that apply the energy equation may suffer from inaccuracy on the conservation of the entropy during the adiabatic processes. For testing and benchmarking of the code, we have reproduced the canonical Moon-forming giant impact \citep{Can04} by comparing with the detailed results provided in \cite{Bar16}.\\

Mars is modeled with a differentiated body with a core-to-mantle mass ratio of 0.3 and a total mass of $6.0 \times 10^{23}$ kg \citep{Cit15,Ros16}. The impactor is set to be an undifferentiated forsterite body with a mass of $1.68 \times 10^{22}$ kg. The total number of SPH particles in the simulation is $N=3 \times 10^5$ or $3 \times 10^6$. The initial spherical body is created by distributing the SPH particles in a 3D lattice (face-centered cubic) and then by performing SPH calculation in isolation until the velocity of particles is much smaller than the escape velocity of the sphere. Initially, the entropy of Mars and the impactor is set to be 2000 J K$^{-1}$ kg$^{-1}$, which corresponds to 680 K at the surface of Mars.\\

Previous simulations of Martian moon-forming disk impacts \citep{Cit15, Ros16} are not sufficient to resolve the disk structure ($N \sim 10^5$). They also used the Tillotson EOS, which has been widely applied to giant impact simulations, but this EOS cannot correctly treat vaporization, and does not provide information about temperature and entropy. Thus, in this work, to compare our results to these previous works, we use a larger number of SPH particles ($N \sim 3 \times10^6$) with the same impact conditions as those in previous works and more sophisticated EOS (M-ANEOS) that can deal with phase change, and temperature and entropy changes. We assume an impact angle of 45 degrees and an impact velocity of 1.4 times the mutual escape velocity ($\sim 6$ km s$^{-1}$) which can form the Borealis basin \citep{Mar08}. Note that, however, the outcome of collision strongly depends on the impact conditions. Thus, future work would require systematic investigation of the disk structure with respect to different impact conditions.

\section{Results}
\subsection{Disk structures and their material compositions}
\begin{figure}[ht!]
\plotone{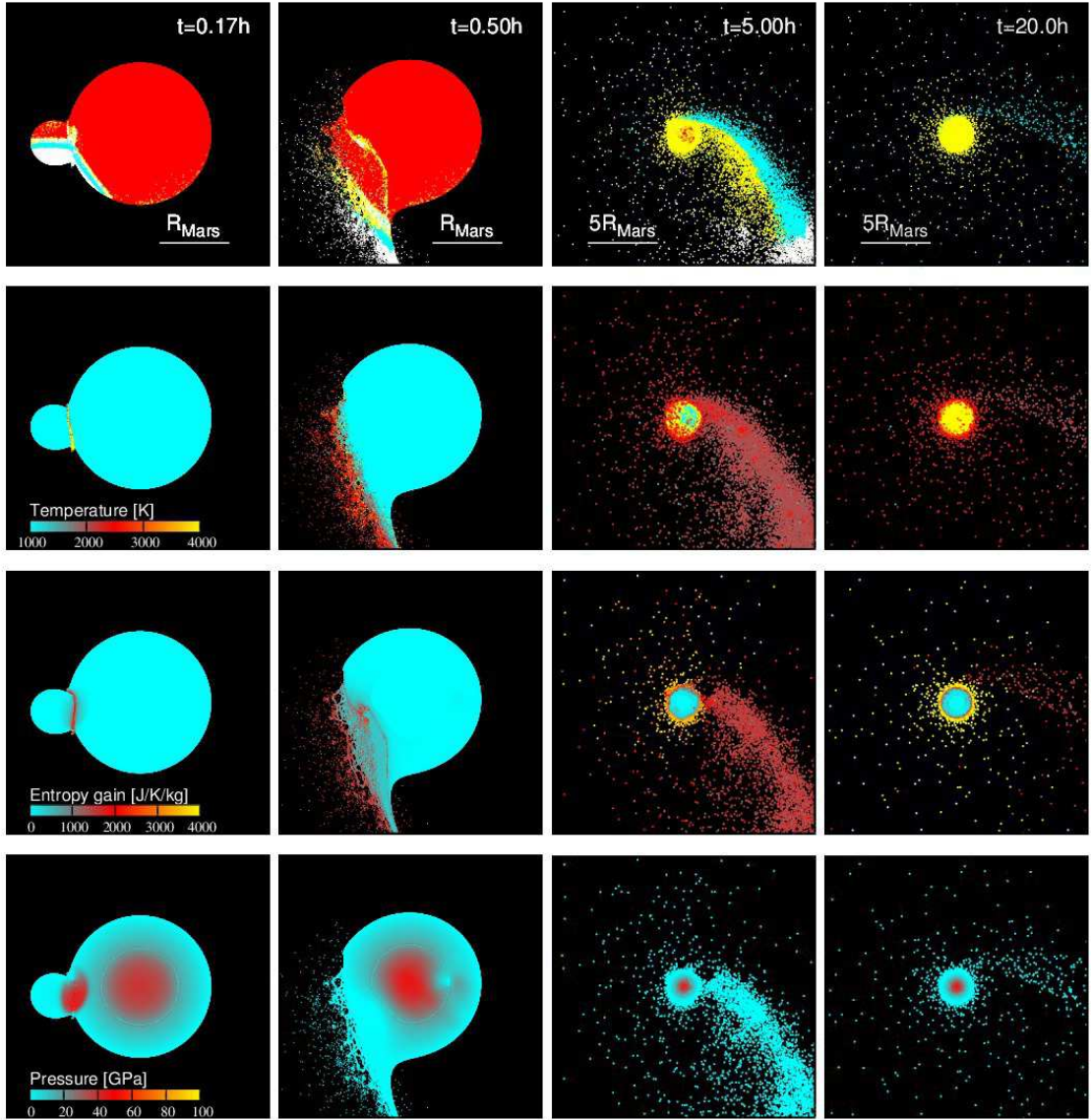}
\caption{Snapshots of the Martian moon-forming impact simulation taken at different epochs ($N=3 \times 10^6$). Each column corresponds to the same epoch. The width of the panel is 10,000 km and 60,000 km for the left two panels and right two panels, respectively. In the top panels, red, yellow, white and cyan points represent Mars, those that fall on Mars, those that escape from Mars, and disk particles, respectively and they are over-plotted in this order. In the second row panels, the color represents temperature (in Kelvin). When the temperature of a particle is below 1000 K or above 4000 K, it is plotted by cyan or yellow, respectively. In the third row panels, the color represents the entropy gain from the initial value of 2000 J K$^{-1}$ kg$^{-1}$. When entropy gain is above 4000 J K$^{-1}$ kg$^{-1}$, it is plotted yellow. In the bottom panels, the color represents the pressure [GPa]. In the third row and bottom panels, only $z<0$ is plotted where $z=0$ is the equatorial plane of Mars. In the top panels, the time $t$ is given in hours. An animation of our canonical giant impact on Mars is available online. In the animation, the color represents temperature (the same as the second raw panels of Figure 1) and the core particles are overplotted by brown. The width of the animation is fixed 60,000 km.}
\label{snapshots}
\end{figure}

As is also shown in the previous works \citep{Cit15, Ros16}, the impact produces the disk around Mars (Figure \ref{snapshots}). From each snapshot of the simulation, particles are systematically sorted into either disk particles, those that consist of Mars, those that fall on Mars, or those that escape from the gravity field of Mars. In order to do so, we calculate the equivalent circular orbital radius $a_{\rm eq}=a(1-e^2)\cos(i)^2$, where $a$, $e$ and $i$ are the semi-major axis, eccentricity and inclination of a particle, respectively. The surface of Mars is determined where the density contour becomes $1,000$ kg m$^{-3}$. When a particle is not gravitationally bound to Mars, it is categorized as an escaping particle. When a particle is gravitationally bound and $a_{\rm eq}>R_{\rm Mars}$, it is categorized as a disk particle. Otherwise, it is categorized as a colliding particle onto Mars. In this work, we use the above criteria to describe the disk particles, so that we can easily compare our results to the previous papers. However, if we take a more restrictive criteria such as the pericenter distance of the particle should be larger than the radius of Mars as $a_{\rm peri} = a(1-e)>R_{\rm Mars}$, the disk mass decreases by a factor of 5. However, these particles hit Mars’ surface with an impact velocity larger than a few km s$^{-1}$ (Figure \ref{orbits}) and thus further erosion of Mars’ surface may occur.\\

\begin{figure}[ht!]
\epsscale{0.6}
\plotone{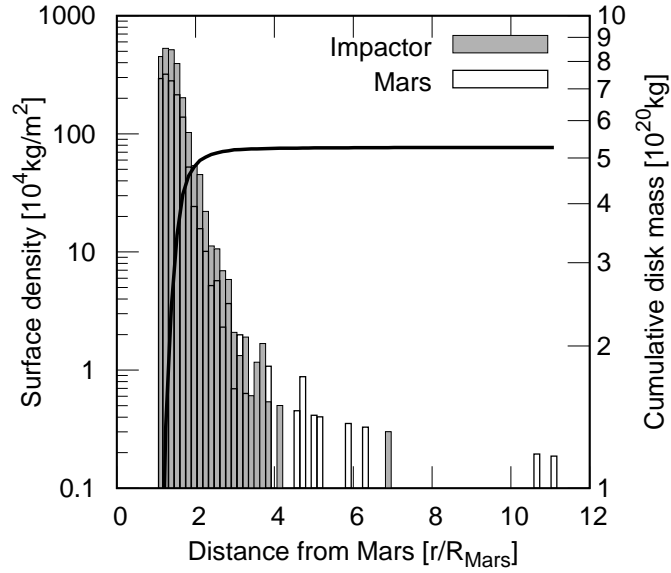}
\caption{Surface density (boxes) and the cumulative mass (solid line) of the disk depending on the distance to Mars ($N=3 \times 10^6$ with an impact angle of $\theta=45$ degrees). The transparent white boxes represent the contribution from Mars material and the gray boxes represent that from the impactor.}
\label{sigma}
\end{figure}

For high-resolution simulation ($N=3 \times 10^6$) with an impact angle of $\theta=45$ degrees, about $2600$ particles are identified as disk material. Most of the disk material is located within the Roche limit, but a small amount of the disk mass is located beyond the Roche limit and even beyond the synchronous orbital radius ($\sim 6R_{\rm Mars}$, Figure \ref{sigma}), which is consistent with previous studies (Fig.1 in \cite{Ros16}). The total disk mass at the steady state is $M_{\rm tot}=5.34 \times 10^{20}$ kg within which the mass that originated from Mars is $37\%$ and the mass that originated from the impactor is $63\%$, respectively. Only looking at the outer disk (beyond 4$R_{\rm Mars}$), we have a total of 45 particles and $\sim 67\%$ of the material comes from Mars and the rest from the impactor, although the detailed distribution of SPH particles in the outer disk is not fully resolved.\\

\begin{figure}[ht!]
\plotone{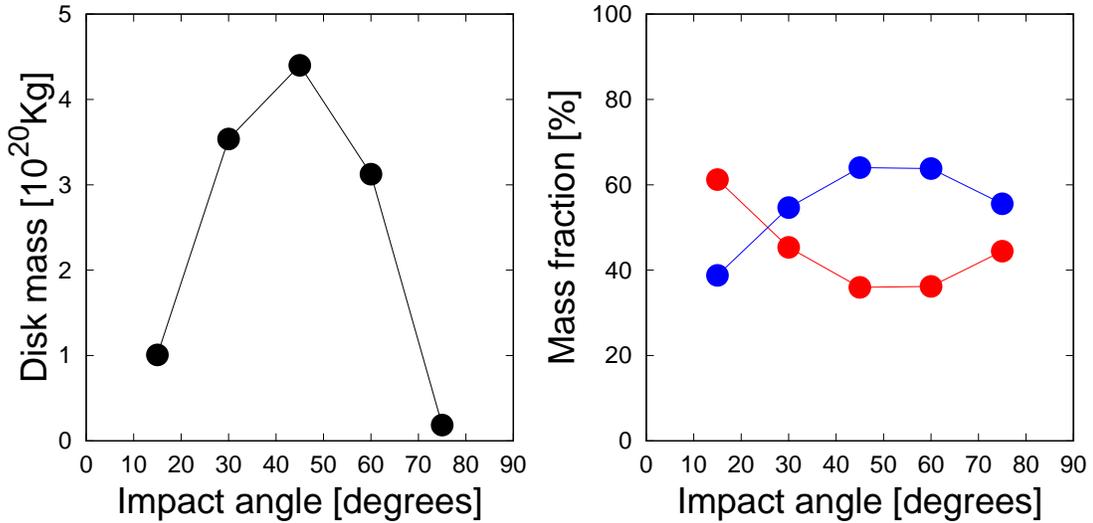}
\caption{Results of SPH simulations ($N=3 \times10^{5}$). Left panel shows disk mass at different impact angles. Right panel shows disk mass fraction that originated from Mars (red points) or from the impactor (blue points) at different impact angles.}
\label{angle}
\end{figure}

We also performed several runs with the same impact conditions but with $N=3 \times 10^5$ and at different impact angles from $15$ to $75$ degrees with 15-degree intervals (Figure \ref{angle}). The resulting disk mass becomes maximum at an impact angle of 45 degrees and the provenance of material from Mars varies between $35$ and $65$\% at different impact angles. Thus, if accretion scenarios \citep{Ros16, Hes17} are correct, both Phobos and Deimos are expected to be a mixture of both Martian and impactor materials.\\

\begin{figure}[h!]
\epsscale{0.6}
\plotone{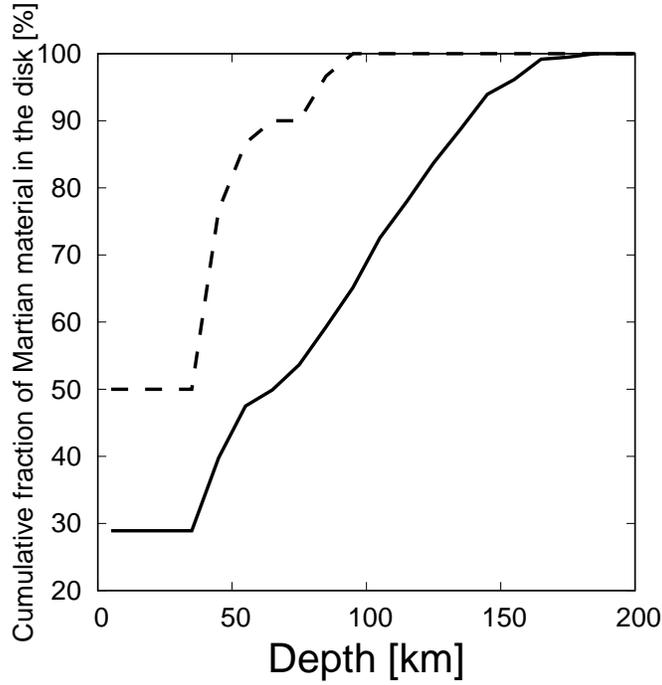}
\caption{Cumulative fraction of disk particles that originated from Mars against their original depth from the surface of Mars ($N=3 \times 10^6$ with an impact angle of $\theta=45$ degrees). Black line represents the case where all disk particles are considered. Dashed line represents the case where only particles whose equatorial circular orbital radius is beyond 4$R_{\rm Mars}$. We take the radius of Mars (before the impact) as 3228 km from our SPH simulation. }
\label{depth}
\end{figure}

Figure \ref{depth} shows the initial location of disk particles inside Mars (as a function of depth). About 30\% of the Martian material in the disk comes from the surface of Mars (down to 10km below the surface). Most of the Martian materials in the disk originated from 50 to 150 km depth (see also Figure \ref{snapshots} top panel cyan color). Considering only the outer disk (beyond 4$R_{\rm Mars}$), the fraction of disk's Martian material that comes from near Mars’ surface increases ($\sim 50\%$), but a large amount of the material still comes from 50 to 150 km in depth (see dashed curves in Figure \ref{depth}). Although the thickness of the crust of early Mars at the time of the giant impact is not well constrained, a significant fraction of the disk’s material comes from the Martian mantle. This is because the crust thickness of today's Mars is estimated to be about 50 km on average \citep{Zub00}, and crust grows with time. Thus, in a framework of the giant impact origin of the Martian moons \citep{Ros16, Hes17}, Phobos and Deimos should contain early Martian mantle material as well.\\

\subsection{Thermodynamic properties of the disk}
\begin{figure}[ht!]
\plotone{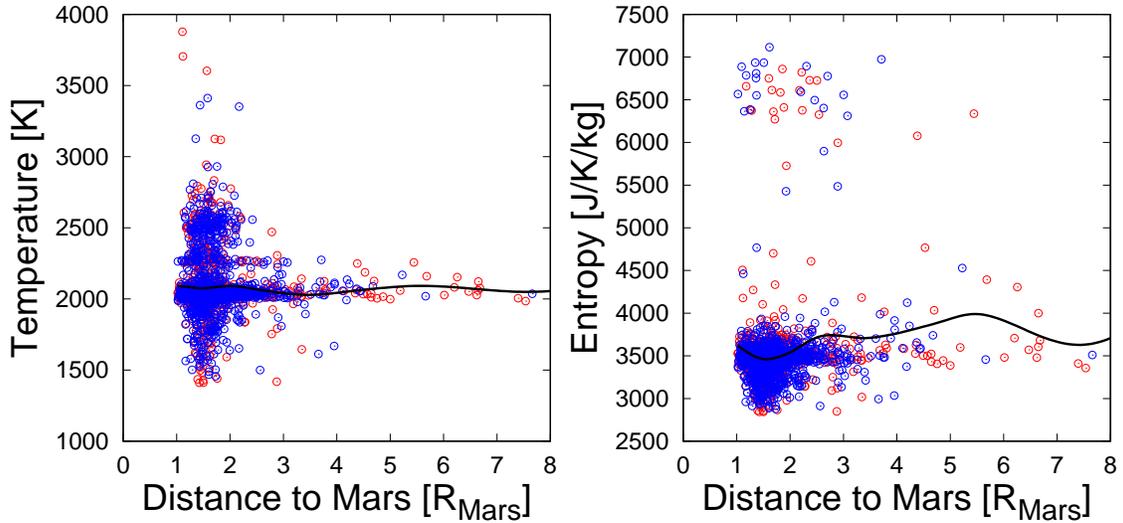}
\caption{Thermodynamic properties of Martian moon-forming disk just after the impact against their expected circular orbital radius ($t=20$h in the case of $N=3 \times 10^6$ with an impact angle of $\theta=45$ degrees). Red points show disk particles that originated from Mars and blue points show those from the impactor. The solid line represents the averaged values of nearby particles. Left and right panels show the temperature and entropy of the disk particles, respectively.}
\label{thermo}
\end{figure}

Figure \ref{thermo} shows the thermodynamic properties of the disk just after the giant impact (temperature, entropy). The disk is relatively uniformly heated up with an entropy increase of 1500 J K$^{-1}$ kg$^{-1}$ (the resulting entropy $\sim 3500$ J K$^{-1}$ kg$^{-1}$) and the disk temperature is constant at about 2000 K (Figure \ref{thermo}). Compared with the case of the canonical Moon-forming impact  \citep{Can04} whose averaged entropy of the disk of $S_{\rm ave} = 4672$ J K$^{-1}$ kg$^{-1}$ \citep{Nak14} and which is the lowest impact energy case among the other Moon-forming impact models \citep{Cuk12, Can12}, the Martian moon-forming impact is much less energetic with an average disk entropy of $S_{\rm ave}=3500$ J K$^{-1}$ kg$^{-1}$.\\
  
In order to determine the vapor fraction $f_{\rm v}$ ($f_{\rm v}$ takes between $0-1$), we use the following lever rule.
\begin{equation}
	S_{\rm disk} = f_{\rm v}S_{\rm v} + \left(  1- f_{\rm v} \right) S_{\rm l},
\end{equation}
where $S_{\rm disk}, S_{\rm l}$ and $S_{\rm v}$ are the specific entropy of the disk material, liquid, and vapor. From M-ANEOS for forsterite \citep[e.g. Fig.4 in][]{Dav17}, $S_{\rm l}$ and $S_{\rm v}$ are estimated to be $\sim 3100$ J K$^{-1}$ kg$^{-1}$  and $11000$ J K$^{-1}$ kg$^{-1}$ at 2000 K, respectively. Since $S_{\rm disk}=3500$ J K$^{-1}$ kg$^{-1}$, the vapor fraction of the disk material is calculated to be less than 5\%.\\

The current M-AMEOS code does not distinguish the solid and liquid phases. For the melting criterion, we use critical entropy gain during the impact as $\Delta S_{\rm melt} \sim 623$ J K$^{-1}$ kg$^{-1}$ \citep{Sti05} as is also used in the previous papers \citep[e.g.][]{Nak15}. In our calculation, the entropy increase is about 1500 J K$^{-1}$ kg$^{-1}$ and thus the disk material is mostly in liquid phase. \cite{Nak17} independently carried out similar simulations, and they showed that the temperature of the disk produced by a giant impact onto Mars becomes 2000 K, which is almost consistent with our results showed here.\\

\subsection{Typical size of disk materials just after the giant impact} 
\label{sec:size}
Since almost all disk material is molten due to the shock heating during the giant impact, the material ejected from the impact behaves like fluid. The size of ejected liquid droplets $r_{\rm p}$ (disk particle with a mass of $m_{\rm p}$) formed by the giant impact is regulated by their shear velocities during the impact and surface tension of the material \citep[e.g.][]{Mel91}. The local kinematic energy of droplets is
\begin{equation}
	K=\frac{1}{2}m_{\rm p}v_{\rm local}^2,
\end{equation}
where $v_{\rm local}=\dot{v}r_{\rm p}$ is the velocity difference between adjacent droplets with $\dot{v}=dv/dr \sim v_{\rm imp}/r_{\rm imp}$. Here, $r_{\rm imp} \sim1000$ km is the radius of the impactor. The surface tension of the silicate melt is $\sigma \sim 0.3$ N m$^{-1}$ \citep{Wal81}. Thus, the balance between these two terms gives us 
\begin{equation}
	r_{\rm p} \sim \left(  \frac{\sigma}{\rho \dot{v}^2} \right)^{1/3}.
\end{equation}

Assuming $\rho = 2500$ kg m$^{-3}$ and using the value of $v_{\rm imp}=6$ km s$^{-1}$, we can calculate the typical size of droplets $r_{\rm p} \sim1.5$ m.\\

\subsection{Cooling timescale of the disk}
Just after the giant impact, the disk particles form an arm-like structure (see Figure \ref{snapshots}). The optical depth is written as $\tau=\Sigma_{i} \sigma_{i}/S$ where $\Sigma_{i} \sigma_{i}$ is the total cross section of particles and $S$ is the local area where particles are distributed. $\Sigma_{i} \sigma_{i}$ is calculated as $\Sigma_{i} \sigma_{i}=N_{\rm tot} \pi r_{\rm p}^2$ where $N_{\rm tot}=M_{\rm disk}/m_{\rm p}$. The initial arm-like structure extends as seen in the fourth quadrant of Figure \ref{snapshots} top panel $t=5$ and $20$h (At $t=5$h, the arm is still expanding). As found in the simulation, the length of the arm can reach about $1 \times 10^8$m. So, $S = 1 \times 10^8$ m $\times$ $1 \times 10^8$ m = 1$ \times 10^{16}$ m$^2$. Assuming a particle has a radius of $1.5$m and a density of $2500$ kg m$^{-3}$, we find $\tau \sim 10$. Thus, we simply assume that the droplets individually cool down with time by radiative emission. Assuming the droplets are black-body, the energy emitted by time unit is
\begin{equation}
	dE/dt = 4 \pi r_{\rm p}^2 \sigma_{\rm SB} T^4,
\end{equation}
where $\sigma_{\rm SB}$ is the Stefan–Boltzmann constant. Thus, the cooling time is 
\begin{equation}
	t_{\rm cool} = \Delta E/(dE/dt),
\end{equation}
with energy change 
\begin{equation}
	\Delta E = m_{\rm p} C_{\rm J} \Delta T,
\end{equation}
where $C_{\rm J}=1000$ J K$^{-1}$ kg$^{-1}$ is the typical specific heat capacity of silicate material. Assuming that a droplet cools down until its solidification temperature of about $1500$ K with $\Delta T=500$ K from 2000 K to 1500 K after the impact, the cooling timescale of meter-sized droplets is calculated to be about $t_{\rm cool} \sim 11$ min. The initial orbital period of the particles just after the impact is about $2$ days assuming the semi-major axis is $10R_{\rm Mars}$ (see also Figure \ref{orbits}). Thus, the droplets solidify soon after the impact and before they orbit around Mars.\\

\subsection{Collisional cascade of the debris}
\begin{figure}[ht!]
\plotone{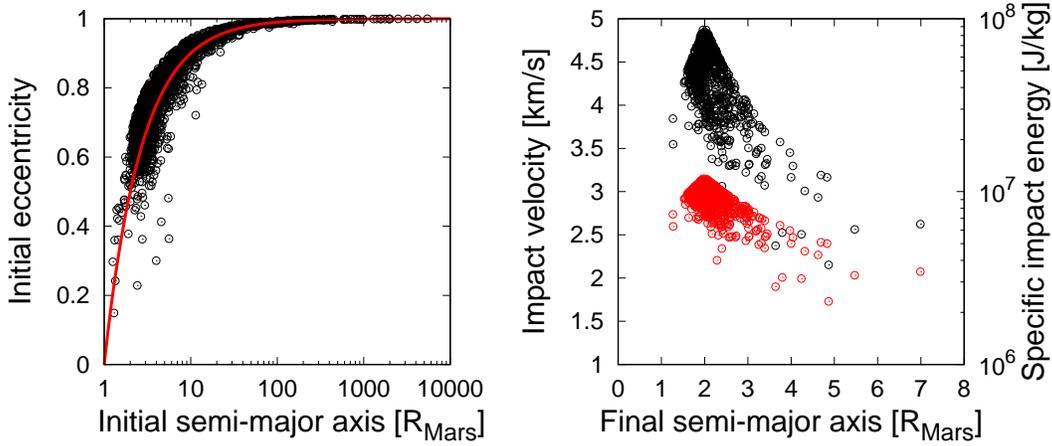}
\caption{Left panel shows eccentricity of particles just after the impact against their semi-major axes ($N=3 \times 10^6$ with an impact angle of $\theta=45$ degrees). The red line represents where pericenter distance equals the Mars radius (3300km). Right panel shows collisional velocity among particles (black dots) and the corresponding specific impact energy ($\frac{1}{2}v_{\rm peri}^2$) assuming the impact occurs at pericenter (red dots) against their circular equivalent orbital radius. In the right panel, only particles whose pericenters are larger than Mars' radius are plotted.}
\label{orbits}
\end{figure}

These solid particles orbit around Mars. Just after the impact, the particles have large eccentricities ($e \sim 0.8$) and slightly different semi-major axes, forming an eccentric ringlet-like structure (Figure \ref{orbits} left panel). As shown in \cite{Hyo17} (their equation 16), the timescale for collision among these particles at this stage can be written as 
\begin{equation}
	t_{\rm coll} = 2\pi a/\left( \frac{3}{2} f r_{\rm p}\Omega \right),
\end{equation}
where $\Omega$ is the Keplerian orbital frequency and $f$ is factor of unity. Using $a=10R_{\rm Mars}$, we calculate $t_{\rm coll} \sim (9/f) \times 10^4$ years.\\

In constrast, the precession rate of the argument of pericenter $\omega$ and the longitude of ascending node $\Omega$ around Mars is much shorter than the above collision timescale as estimated below. The precession rates due to flatness of Mars ($J_{2}= 1.96 \times 10^{-3}$) can be estimated by \citep{Kau66, Hyo17}
\begin{equation}
	\dot{\omega} = \frac{3n}{\left( 1-e^2 \right)^2} \left( \frac{R_{\rm Mars}}{a} \right)^2 \left( 1-\frac{5}{4}\sin^2(i) \right)J_2
\end{equation}
\begin{equation}
	\dot{\Omega} =  -\frac{3n\cos(i)}{2\left( 1-e^2 \right)^2} \left( \frac{R_{\rm Mars}}{a} \right)^2 J_2,
\end{equation}
where $n$ is the orbital mean motion of particles. Assuming, $a=10R_{\rm Mars}$, $e=0.8$, $i=45$ degrees, we calculate the timescale of precession $\tau_{\omega}=2\pi/\dot{\omega} \sim 35$ years and $\tau_{\Omega}=2\pi/\dot{\Omega} \sim 37$ years, respectively. Thus, precession occurs quickly and forms a torus-like structure before they collide.\\

Once the orbital direction is randomized, the orbits of the particles cross and collision velocities become comparable to local Keplerian velocities at around pericenter \citep{Hyo17}. Assuming particles collide at their pericenter, the collision velocity is about : 
\begin{equation}
	v_{\rm peri} = na \sqrt{ \frac{1+e}{1-e} }.
\end{equation}
The collisional velocity is between $1-5$ km s$^{-1}$ and the specific impact energy becomes $\sim 10^6$-$10^7$ J kg$^{-1}$ (Figure \ref{orbits}). Some fraction of this impact energy is used for temperature increase, melting, and vaporization. Since the latent heat of melting for silicate ($\sim 6 \times10^5$ J kg$^{-1}$ for forsterite in \cite{Nav89}) is smaller than or comparable to the specific impact energy, we expect that most of the colliding particles melt again. On the other hand, since the latent heat of vaporization for silicate ($\sim 10^7$ J kg$^{-1}$ in \cite{Pah07}) is larger than the specific impact energy, we expect that vaporization due to the collision cascade of particles in the disk does not effectively occur. When melting occurs, using the same arguments in Section \ref{sec:size} and assuming a collisional velocity of $3-5$ km s$^{-1}$, meter-sized particles become $\sim 100$ micro-meter sized droplets, followed by quick solidification.\\

During the initial particle-particle collisions, the kinematic energy is damped and eccentricity decreases. Thus, successive collision becomes much less energetic and further melting impacts are less likely to occur. Without melting, it is not simple to determine the typical size of particles after the particle-particle collsion. Therefore, the size distribution of the final building blocks of Phobos and Deimos are expected to be very wide, from meter-sized particles fragmented during the giant impact, to 100 $\mu$m particles fragmented during subsequent collision cascade, down to 0.1$\mu$m particles condensed from silicate vapor \citep{Ron16} produced during the giant impact.\\

The surface features such as spectral properties are related to its grain size \citep{Pie16}. We consider the case where two different sized solid particles, $r_{\rm p1}$ and $r_{\rm p2}$, are well-mixed ($r_{\rm p1} < r_{\rm p2}$) with their mass fraction of $r_{\rm p1}$ sized particle $f$ and that of $r_{\rm p2}$ sized particles $1-f$. The total surface area of the larger particles can be written as  
\begin{equation}
	S_{\rm 2,tot} = N_{\rm 2} \times 4 \pi r_{\rm p2}^2,
\end{equation}
where $N_{\rm 2} = \frac{(1-f) M_{\rm disk}}{m_{\rm 2}}$ is the total number of $r_{\rm p2}$ sized particles. The sum of the cross section of the smaller particles can be written as 
\begin{equation}
	\sigma_{\rm 1,tot} = N_{\rm 1} \times  \pi r_{\rm p1}^2,
\end{equation}
where $N_{\rm 1} = \frac{f M_{\rm disk}}{m_{\rm 1}}$ is the total number of $r_{\rm p1}$ sized particles. The ratio between the total surface area of the larger particles and the total cross section of the smaller particles can be written as 
\begin{equation}
	\frac{S_{\rm 2,tot}}{\sigma_{\rm 1,tot}} = 4\left( \frac{1-f}{f} \right) \left( \frac{r_{\rm p1}}{r_{\rm p2}} \right).
\end{equation}
In our case, if we consider $r_{\rm p1}=0.1$ $\mu$m with $f=0.05$ and $r_{\rm p2}=100$ $\mu$m with $f=0.95$, we can calculate $\frac{S_{\rm 2,tot}}{\sigma_{\rm 1,tot}} \sim 0.08$. This means that all surfaces of the large particle can be covered by $0.1$ $\mu$m sized small particles. Thus, although the building blocks of Phobos and Deimos are the mixture of different sized particles, 0.1$\mu$m particles condensed from vapor are likely to be representative of the surface properties.

 \section{Summary \& Discussion}
The origin and dynamical evolution of the two small Martian moons, Phobos and Deimos, are intensely debated. \cite{Cra94,Cra11} have proposed that Phobos and Deimos may form within a debris disk that formed by a giant impact. Recent works have shown that they can be formed by accretion within a debris disk that is formed by a giant impact \citep{Ros16,Hes17}. These previous works mostly focused on explaining the mass and orbital properties of the two moons, but the detailed analysis of the disk properties (that is, the building blocks of the Martian moons) have scarcely been done. In this work, we performed high-resolution SPH simulations and investigated the detailed structure and thermodynamic properties of an impact-induced Martian moon-forming disk.\\

We have considered the canonical Martian moon-forming impact whose impact energy is $3 \times 10^{29}$ J and impact angle is 45 degrees  \citep{Cit15,Ros16}, which can form the Borealis basin \citep{Mar08}. We showed that the resulting debris disk is expected to experience two stages of thermodynamic evolution.\\

1. As a direct consequence of the giant impact, the disk material is heated up to 2000K with an entropy increase of 1500 J K$^{-1}$ kg$^{-1}$ and thus most of the disk material is in liquid phase, but a small amount of material (less than 5wt\%) is vaporized. The size of the liquid droplets is regulated by shear velocity and surface tension during the ejection phase on impact and they become meter-sized droplets followed by quick solidification to solid particles (Section 3.3).\\

2. The solid particles initially have large eccentricities. Then, they experience precession around Mars due to mainly Mars’ $J_2$ term. After the orbits are randomized, their orbits cross and collide with an impact velocity of $1-5$ km s$^{-1}$. These high-velocity collisions can partly result in melting and form at minimum $\sim 100$ $\mu$m-sized particles from meter-sized particles (Section 3.5).\\

\cite{Ron16} have shown that if the building blocks of Phobos and Deimos are formed from a magma disk, the size of particles becomes too coarse ($0.1-1$ mm) to explain the observed spectral features of their surfaces. In contrast, if they experience gas-to-solid condensation, the size of the particles is expected to be smaller than 2 microns and explains the observed similarities between the spectral properties of D-type asteroids and that of the Martian moons’ surfaces \citep{Ron16}. In this study, we have shown that a small amount of disk material is expected to vaporize: less than 5\% during the initial giant impact phase (Section 3.2). The gas-to-solid condensation of this vapor material can form $\sim 0.1$ micro-meter sized particles \citep[e.g.][]{Ron16}. The total mass of this $0.1$ $\mu$m sized particles is small but their total cross section is larger than that of other larger particles. Thus, the mixing of these fine grains and other larger particles (100$\mu$m - meter in size) may be able to reconcile with the observed surface properties of Phobos and Deimos. Therefore, the above two stages of evolution may be a successful path to form Phobos and Deimos with a good reproduction of both their physical and orbital characteristics. In addition, the micro-meteoritic bombardement tends to reduce the size of the grains at the surface, hence it may be able to reinforce the effect of microns-size particles on the reflectance spectra as argued by \cite{Ron16}.\\

If some water is contained inside the impactor or the surface of Mars, water is expected to be ejected and vaporized during the giant impact process and may form a water vapor cloud around the disk. Regardless of whether or not this vapor cloud escapes from the disk, there would be a chance for the water vapor to dissolve into the disk particles of silicate melt. This indicates that the building blocks of Phobos and Deimos contain small amounts of water as hydrous minerals. Note that, however, space weathering and implantation of micrometeorites (contamination by external materials), may change the surface environment after the formation of these two airless small bodies. We leave this matter for future works.\\

If the \cite{Ros16} scenario is correct, Phobos and Deimos accrete in an extended outer disk (4-7$R_{\rm Mars}$). In contrast, following the arguments of \cite{Hes17}, Phobos is the final outcome of the cycle of accretion and destruction of a moon that formed from the spreading of the Roche-interior disk. So, the radial provenance of the building blocks of Martian moons depends on the models. We investigated the provenance of disk material: either from Mars or the impactor. We find that in the case of a 45 degrees impact, about 40\% of the total disk mass originated from Mars and the rest comes from the impactor. Considering only the outer disk (beyond 4$R_{\rm Mars}$), $\sim 70$\% comes from Mars. Changing the impact angle results in a different mixing rate, but the whole disk contains Martian material of at least 35wt\% (Figure \ref{angle}). In addition, the Martian material is largely expected to come from the Martian mantle ($50-150$ km in depth below the surface; Figure \ref{depth}). It is generally difficult to obtain such deep material by spacecraft activities. In contrast, a giant impact is a natural drilling event that provides us with an opportunity to get Martian mantle material. Therefore, Japan's upcoming sample return mission (MMX) may have the chance to obtain such Martian samples if either accretion scenario \citep{Ros16,Hes17} is correct. Our model predictions presented in this paper would be tested by this Japanese sample return (MMX) mission.\\

\acknowledgments
We acknowledge the financial supports of the UnivEarthS Labex programme at Sorbonne Paris Cit{\'e} (ANR-10-LABX-0023 and ANR-11-IDEX-0005-02), JSPS Grants-in-Aid for JSPS Fellows (17J01269), JSPS Grants-in-Aid for Scientific Research (15K13562), and the JSPS-MAEDI bilateral joint research project (SAKURA program). This work was also supported by Université Paris Diderot and by a Campus Spatial grant. Numerical computations were partly performed on the S-CAPAD platform, IPGP, France.

\vspace{5mm}
\software{GADGET-2 (Springel 2005)}

\end{document}